# Ionic-Radius Mismatch as a Structural Lever for Tuning Phase Transitions and Luminescent Thermometry

**A. Javaid[1], M. Szymczak[1*], A. Sieradzki[2], L. Marciniak[1*]**

[1] Institute of Low Temperature and Structure Research, Polish Academy of Sciences,

Okólna 2, 50-422 Wrocław, Poland

[2] Department of Experimental Physics, Wrocław University of Technology, Wybrzeże Wyspiańskiego 27, 50-370 Wrocław, Poland

*corresponding author: m.szymczak@intibs.pl , l.marciniak@intibs.pl

*KEYWORDS luminescent thermometry, phase transition, high sensitivity, modulation, thermometric performance*

## Abstract

The widespread implementation of luminescence thermometers requires a comprehensive understanding of the structural factors governing their thermometric performance. Establishing such structure-property relationships is essential for the rational design of sensing materials with application-tailored characteristics. This is particularly relevant for phase-transition-based luminescence thermometers, which offer exceptionally high relative sensitivities. The systematic analysis of $K_3Lu(PO_4)_2:Eu^{3+}$ demonstrates that introducing co-dopant ions with a controlled ionic-radius mismatch provides an effective strategy for tailoring phase-transition characteristics. This approach enables both the phase-transition temperature and thermal operating range to be controlled. Specifically, the transition temperature shifts from 210 K for $K_3Lu(PO_4)_2:Eu^{3+}$ to 310 K for $K_3Lu(PO_4)_2:Eu^{3+}$,10%$La^{3+}$, while the operating range broadens from 30 to 60 K. Importantly, linear correlations between the ionic-radius mismatch parameter, Ω, and the phase-transition temperature, enthalpy, and entropy provide a quantitative framework for controlling the thermodynamics of the transition through compositional engineering. Beyond luminescence thermometry, these relationships establish a general strategy for designing materials exhibiting first-order phase transitions with tailored thermodynamic characteristics, opening opportunities for their optimization across a broad range of functional applications.

**Introduction**

The spectroscopic properties of trivalent lanthanide ions are generally considered to exhibit only limited sensitivity to changes in the local crystallographic environment they occupy [1–4]. This is particularly true when compared with the pronounced effects observed in phosphors doped with transition metal ions[5,6]. Nevertheless, a detailed analysis of lanthanide spectroscopy reveals that variations in local site symmetry may give rise to a number of significant changes in both the emission spectra and the radiative relaxation dynamics[7–9]. These changes are manifested primarily through variations in the intensities of hypersensitive electronic transitions and in the number of Stark components into which the electronic energy levels are split, ultimately affecting the number and distribution of emission lines observed in the luminescence spectra[10–12].

Although these spectroscopic features provide valuable information for fundamental studies, which is why certain lanthanide ions, such as $Eu^{3+}$, are commonly regarded as luminescent structural probes, they can also be successfully exploited for sensing applications [7,8,10–12]. One of the most attractive examples is the combination of lanthanide luminescence with host materials exhibiting reversible thermally induced first-order phase transitions for temperature sensing. In such systems, abrupt changes in the luminescence properties of the lanthanide ions accompanying the structural phase transition are utilized to construct both ratiometric and lifetime-based luminescence thermometers[13–19]. Owing to the discontinuous nature of these structural transformations, the resulting thermal sensitivities are often several times higher than those reported for conventional lanthanide-based luminescence thermometers relying on thermally coupled excited states or interionic energy transfer processes[20–23].

Despite these Ωntages, phase-transition-based luminescence thermometers suffer from a relatively narrow thermal operating range, which is intrinsically limited to the temperature interval over which the structural phase transition occurs[14,15,17–19,24–31]. Consequently, shifting the operating temperature range generally requires replacing the host material with another compound possessing a different phase transition temperature. However, such a strategy is frequently accompanied by changes in the mechanical, thermal, and chemical stability of the luminescent thermometer, which may compromise its practical applicability.

To overcome this limitation, an alternative approach has recently been proposed, in which the phase transition temperature, and consequently the operating temperature range, is tuned through the incorporation of optically inactive dopant ions whose ionic radii differ from those of the host cations[14,15,17,28,31]. This strategy enables the operating temperature of the luminescent thermometer to be tailored to the requirements of a specific application while preserving nearly identical mechanical and chemical properties of the host material. However, although this approach effectively controls the position of the operating temperature range, it does not provide a means to regulate its width.

In the present work, we therefore investigate the influence of optically inactive co-dopants on not only the phase transition temperature but also the thermal operating range of phase-transition-based luminescence thermometers. To this end, the spectroscopic properties of $K_3Lu(PO_4)_2:Eu^{3+}$ were systematically investigated over a wide temperature range as a model luminescent thermometer. The obtained results were compared with those of analogous phosphors co-doped with $Sc^{3+}$, $La^{3+}$, and $Y^{3+}$ ions. The performed analysis demonstrates that the incorporation of these optically inactive dopants enables effective control over the thermal operating range of the luminescent thermometer. Furthermore, the observed behavior is compared with that previously reported for phase-transition-based luminescence thermometers based on $LiYO_2:Eu^{3+},Na^+$ and $LaGaO_3:Eu^{3+},Sc^{3+}$, providing broader insight into the role of optically inactive co-doping in engineering the operating characteristics of this emerging class of luminescent thermometers.

## Experimental Section

### *Synthesis*

A powered samples of $K_3Lu_{0.9}M_{0.1}(PO_4)_2$:1%$Eu^{3+}$ (where M= $Sc^{3+}$, $Y^{3+}$ and $La^{3+}$) were synthesized using a conventional high temperature solid-state reaction method. $K_2CO_3$ (99.9% of purity, Alfa Aesar), $NH_4H_2PO_4$ (99.9% of purity, POL-AURA), $Lu_2O_3$ (99.999% of purity, Stanford Materials Corporation), $Sc_2O_3$ (99.999% of purity, Stanford Materials Corporation), $Y_2O_3$ (99.999% of purity, Stanford Materials Corporation), $La_2O_3$ (99.999% of purity, Stanford Materials Corporation) and $Eu_2O_3$ (99.999 % of purity, Stanford Materials Corporation) were used as starting materials. The raw materials were calculated based on the stoichiometric ratio, precisely weighed and finely ground with few drops of hexane in an agate mortar to achieve a homogeneous mixture. The mixture was subsequently transferred to an alumina crucible and calcined in air at 1573 K for 5 hours (with a heating rate of 10 K $min^{-1}$). The final powders were allowed to cool naturally to the room temperature and then ground again to obtain powder samples for structural and optical characterization.

### *Characterization*

The obtained materials were examined using powder X-ray diffraction technique. Powder diffraction data were obtained in Bragg–Brentano geometry using a PANalytical X'Pert Pro diffractometer using Ni-filtered Cu K$\alpha$ radiation (V=40 kV, I=30 mA).

Calorimetric investigations were carried out using a Mettler Toledo DSC-3 differential scanning calorimeter. The samples were enclosed in aluminum pans with perforated lids and measured under a continuous nitrogen purge. Thermograms were recorded over a temperature range of 120-400 K at a constant heating rate of 5 K min$^{-1}$. To evaluate the excess heat capacity associated with the phase transition, a suitable baseline was subtracted from the experimental heat-flow data.

The excitation spectra were obtained using the FLS1000 Fluorescence Spectrometer from Edinburgh Instruments equipped with 450 W Xenon lamp and R928 photomultiplier tube from Hamamatsu as a detector. Emission spectra were measured using the same system with 980 nm laser diodes as excitation source.

**Results and discussion**

$K_3Lu(PO_4)_2$ can crystallize in two distinct crystal systems: a monoclinic structure with the space group $P2_1/m$ and a trigonal structure with the space group $P3$ (Figure 1a) [32–40]. Notably, the monoclinic form exists in two polymorphic forms, referred to as the $\alpha$ (low temperature phase, LT) and $\beta$ (medium temperature phase, MT) phases. Upon heating, $K_3Lu(PO_4)_2$ undergoes $\alpha \rightarrow \beta$ phase transition, which is accompanied by an increase in structural symmetry and a significant reorganization of the local coordination environment of $Lu^{3+}$ ions. As a result, the coordination number of $Lu^{3+}$ decreases from 7 to 6, leading to a substantially more regular $Lu^{3+}$-$O^{2-}$ polyhedron (Figure 1b). In the $\alpha$ phase, $Lu^{3+}$-$O^{2-}$ bond lengths span a broad range from 2.215 to 2.567 Å, whereas in the $\beta$ phase the $LuO_6$ polyhedra are considerably less distorted, with bond lengths ranging from 2.159 to 2.256 Å. A further increase in temperature induces the $\beta \rightarrow \gamma$ (high temperature phase, HT) phase transition, resulting in the formation of the trigonal phase ($P3$), which possesses the highest symmetry among the three phases. In this phase, $Lu^{3+}$ ions occupy nearly ideal $LuO_6$ octahedra characterized by uniform $Lu^{3+}$-$O^{2-}$ bond lengths of 2.199 Å [32–40]. These pronounced structural transformations strongly affect the local crystal field around the $Lu^{3+}$ ions. Since $Eu^{3+}$ ions substitute for $Lu^{3+}$ in the lattice, changes in the coordination environment, site symmetry, and polyhedral distortion lead to substantial modifications of the spectroscopic properties of $Eu^{3+}$, including its emission spectra, luminescence intensity, and decay kinetics. Consequently, the optical response of $Eu^{3+}$ provides a highly sensitive probe of the phase transitions occurring in $K_3Lu(PO_4)_2$.

A comparison of the XRD patterns of $K_3Lu(PO_4)_2$:1%$Eu^{3+}$ containing different co-dopant ions confirms the formation of a single-phase $K_3Lu(PO_4)_2$ structure in all investigated samples, consistent with the reference pattern reported in ICSD 191224 (Figure 1c). Nevertheless, the incorporation of co-dopants with ionic radii different from that of the host cation $Lu^{3+}$ ($R$ = 0.861 Å) induces subtle modifications of the crystal lattice. In particular, the ionic radii of $Sc^{3+}$ ($R$ = 0.745 Å), $Y^{3+}$ ($R$ = 0.9 Å), and $La^{3+}$ ($R$ = 1.032 Å) differ from that of $Lu^{3+}$, leading to measurable changes in the unit-cell dimensions. As a consequence, slight shifts in the positions of the diffraction peaks are observed depending on the type of co-dopant introduced. This effect is particularly evident for the most intense diffraction reflection, whose position changes from $2\theta$ = 29.74° for $K_3Lu(PO_4)_2$:1%$Eu^{3+}$ to 29.85°, 29.71°, and 29.68° for the $Sc^{3+}$, $La^{3+}$, and $Y^{3+}$-co-doped samples, respectively (Figure 1d). These systematic shifts indicate a contraction of the crystal lattice in the presence of $Sc^{3+}$ ions and an expansion of the unit cell when larger $Y^{3+}$ and $La^{3+}$ ions are incorporated.

The observed structural modifications are consistent with the differences in ionic radii between the host and dopant ions and provide direct evidence that co-doping effectively alters the lattice parameters of $K_3Lu(PO_4)_2$:1%$Eu^{3+}$. Such changes in the crystal structure are expected to influence the local strain environment and may consequently affect the phase-transition behavior and spectroscopic properties of the investigated materials.

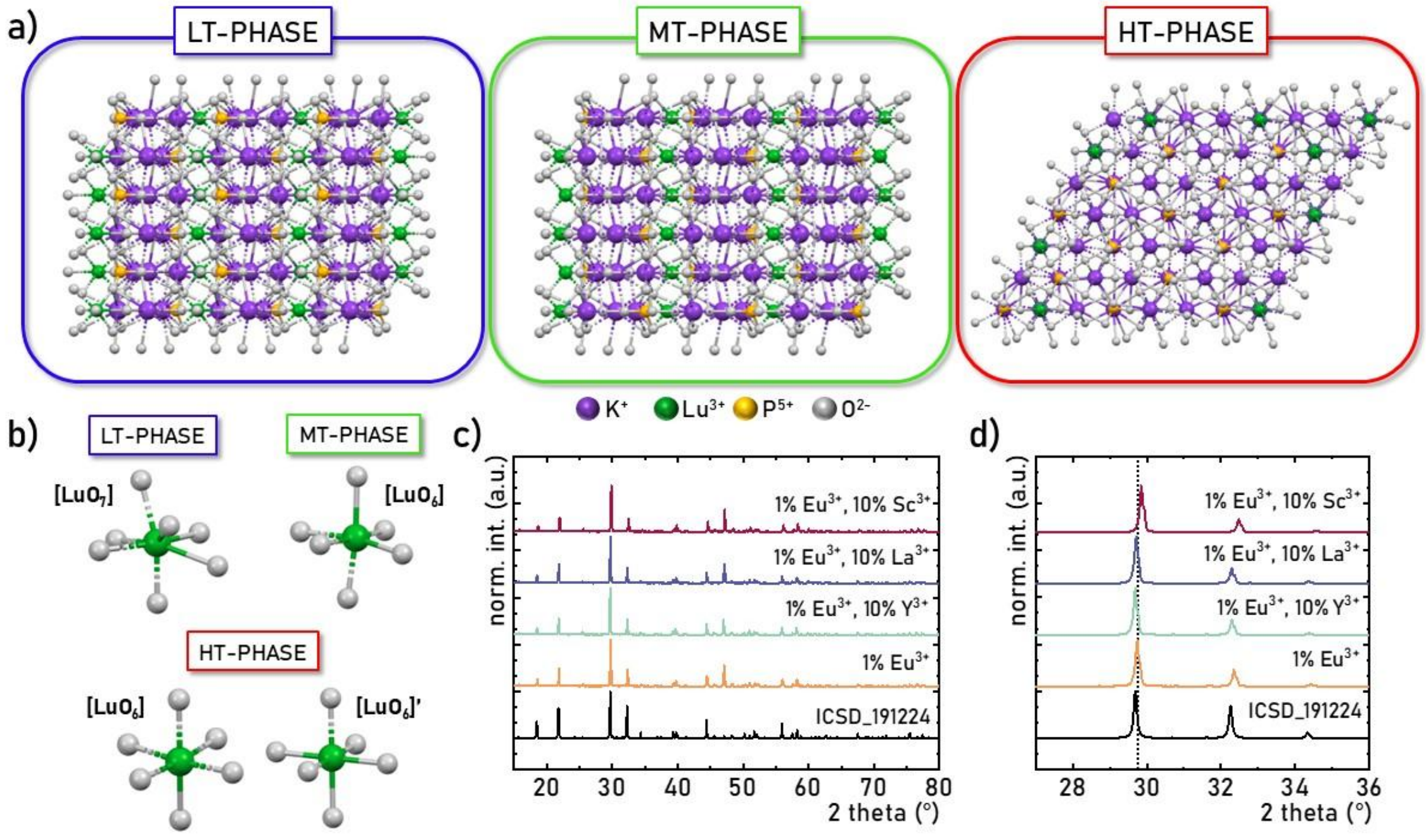

**Figure 1.** Visualization of structures of low temperature (LT), medium temperature (MT) and high temperature (HT) phases of $K_3Lu(PO_4)_2$ – a) and $Lu^{3+}$ ions sites – b) the comparison of the room temperature XRD patterns of $K_3Lu(PO_4)_2$:1%$Eu^{3+}$ with different co-dopants – c); magnified view of the 27-36° 2θ range – d).

The development of luminescence thermometers based on thermally induced structural phase transitions requires the selection of a luminescent ion that exhibits high sensitivity to changes in the local crystal structure. In this context, $Eu^{3+}$ ions represent an ideal choice, as they are widely recognized as luminescent structural probes due to their exceptional sensitivity to variations in the local crystallographic environment[7–12]. This sensitivity originates from several characteristic features of the $Eu^{3+}$ emission spectrum (Figure 2a). The luminescence of $Eu^{3+}$ is dominated by transitions from the $^5D_0$ excited state to the $^7F_J$ multiplets, among which the most intense emission bands are located at approximately 590 nm and 620 nm. These bands correspond to the $^5D_0 \rightarrow {}^7F_1$ magnetic-dipole transition and the $^5D_0 \rightarrow {}^7F_2$ electric-dipole transition, respectively. While the intensity of the magnetic-dipole transition is generally considered to be largely independent of the local site symmetry, the intensity of the electric-dipole transition is highly sensitive to the crystal-field environment. Consequently, structural modifications around $Eu^{3+}$ ions are expected to induce pronounced changes in the relative intensities of these two emission bands. Furthermore, all $^7F_J$ multiplets except $^7F_0$ undergo crystal-field splitting into Stark components. The number of Stark components increases with decreasing point symmetry of the crystallographic site occupied by $Eu^{3+}$ ions and with increasing $J$ value[7]. Therefore, both the number and spectral distribution of Stark lines serve as highly sensitive indicators of structural changes occurring within the host lattice. For these reasons, $K_3Lu(PO_4)_2$:1%$Eu^{3+}$ was doped with $Eu^{3+}$ ions and investigated as a model system (Figure 2 b-c). The emission spectrum of $K_3Lu(PO_4)_2$:1%$Eu^{3+}$ exhibits four prominent bands centered at approximately 590 nm, 620 nm, 660 nm, and 700 nm, corresponding to the $^5D_0 \rightarrow {}^7F_1$, $^5D_0 \rightarrow {}^7F_2$, $^5D_0 \rightarrow {}^7F_3$, and $^5D_0 \rightarrow {}^7F_4$ transitions, respectively. Temperature-dependent emission measurements reveal substantial changes in the spectral shape. The thermal map of the normalized emission spectra clearly indicates the presence of three distinct spectral signatures corresponding to the three crystallographic phases of $K_3Lu(PO_4)_2$:1%$Eu^{3+}$ (Figure 2d-g). Below 200 K, the emission spectrum is characterized by comparable intensities of the $^5D_0 \rightarrow {}^7F_1$ and $^5D_0 \rightarrow {}^7F_2$ transitions. Above 200 K, the intensity of the $^5D_0 \rightarrow {}^7F_2$ band increases relative to that of the magnetic-dipole transition, while the number of Stark components remains essentially unchanged. Upon further heating above approximately 270 K, a pronounced modification of the emission profile occurs. In this temperature range, the $^5D_0 \rightarrow {}^7F_1$ transition becomes significantly more intense than the $^5D_0 \rightarrow {}^7F_2$ transition. Simultaneously, the number of observable Stark components decreases, although this effect is partially obscured by thermal broadening of the emission bands. The reduction in the intensity of the electric-dipole transition may be attributed to an increase in the local symmetry of the crystallographic sites occupied by $Eu^{3+}$ ions accompanying the structural phase transition. To illustrate these structural effects more clearly, a detailed comparison of the MT and HT phase emission spectra was performed separately for each emission band. This analysis demonstrates that the MT → HT phase transition is accompanied by both a reduction in the number of Stark components and a shift in their spectral positions. The effect is particularly evident for the $^5D_0 \rightarrow {}^7F_1$ transition. Owing to the relatively low value of *J*, the number of Stark components associated with this transition remains limited for both phases, thereby minimizing spectral overlap between the MT- and HT-related features and facilitating their discrimination.

An additional parameter sensitive to the structural modifications occurring in $K_3Lu(PO_4)_2:Eu^{3+}$ is the intensity ratio $LIR_1$, defined as:

$$LIR_1 = \frac{\int_{600nm}^{620nm} {}^5D_0 \rightarrow {}^7F_2 \, d\lambda}{\int_{580nm}^{600nm} {}^5D_0 \rightarrow {}^7F_1 \, d\lambda} \quad (1)$$

The value of this parameter changes markedly across the different phases, increasing from 1.52 in the LT phase to 2.16 in the MT phase, followed by a decrease to 0.94 in the HT phase (Figure 2h). Temperature-dependent luminescence studies of $K_3Lu(PO_4)_2:1\%Eu^{3+}$ co-doped with $Y^{3+}$, $Sc^{3+}$, and $La^{3+}$ revealed no substantial alterations in the overall spectral shape (Figure 2i, see also Figure S1-S4). Nevertheless, the introduction of these co-dopant ions modifies the relative intensities of the $^5D_0 \rightarrow {}^7F_1$ and $^5D_0 \rightarrow {}^7F_2$ transitions. As a consequence, $LIR_1$ changes to 2.1 for $Sc^{3+}$, 1.5 for $Y^{3+}$, and 1.8 for $La^{3+}$ (Figure 2j). These observations indicate that the incorporation of co-dopant ions alters the local symmetry around $Eu^{3+}$ ions through lattice strain induced by the ionic-radius mismatch between the host cations and the introduced dopants. As a result, co-doping not only affects the structural phase-transition behavior but also modifies the spectroscopic response of $Eu^{3+}$, further confirming the sensitivity of this ion to subtle changes in the local crystal environment.

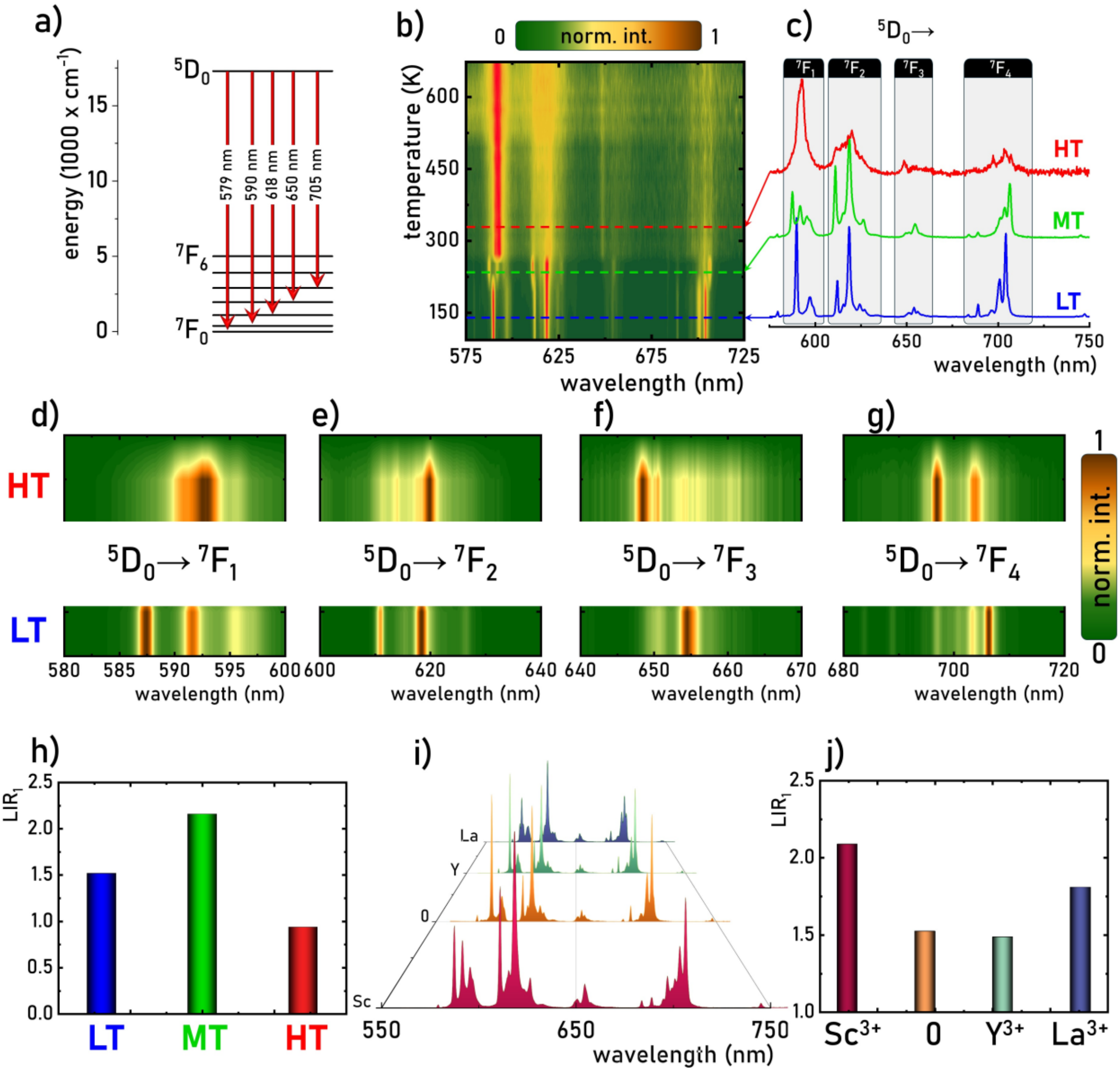

**Figure 2.** Simplified energy levels diagram of $Eu^{3+}$ ions – a); thermal map of normalized emission spectra of $K_3Lu(PO_4)_2$:1%$Eu^{3+}$ - b); representative emission spectra of $K_3Lu(PO_4)_2$:1%$Eu^{3+}$ for low temperature (LT) medium temperature (MT) and high temperature (HT) phases of $K_3Lu(PO_4)_2$:1%$Eu^{3+}$ -c); comparison of luminescence maps of emission spectra of $K_3Lu(PO_4)_2$:1%$Eu^{3+}$ for LT and HT phases in the spectral range corresponding to the $^5D_0\rightarrow{}^7F_1$ – d); $^5D_0\rightarrow{}^7F_2$ – e); $^5D_0\rightarrow{}^7F_3$ – f); $^5D_0\rightarrow{}^7F_4$ – g) electronic transitions; $LIR_1$ for LT and HT phases – h); the comparison of the normalized emission spectra of $K_3Lu(PO_4)_2$:1%$Eu^{3+}$ and co-doped with 10% of $Sc^{3+}$, $Y^{3+}$ and $La^{3+}$ at 93 K – i) and corresponding $LIR_1$ – j).

To gain deeper insight into the influence of 10% $Sc^{3+}$, $La^{3+}$, and $Y^{3+}$ co-doping on the structural phase-transition behavior of $K_3Lu(PO_4)_2$:1%$Eu^{3+}$, temperature-dependent emission spectra were recorded and the corresponding normalized thermal maps are presented in Figure 3a-d. Analysis of these results reveals significant differences in phase stability depending on the nature of the co-dopant ion. For $K_3Lu(PO_4)_2$:1%$Eu^{3+}$, 10%$Sc^{3+}$, only the MT and HT phases are observed within the investigated temperature range, with the MT → HT phase transition occurring at approximately 224 K. In contrast, $K_3Lu(PO_4)_2$:1%$Eu^{3+}$, 10%$Y^{3+}$ looks like exhibits only the LT and HT phases, with a direct phase transition between them observed at approximately 285 K. However, a careful analysis of the thermal dependence of emission spectra of this phosphor indicates that MT phases occurs only in a very narrow thermal range for this compound between 270 - 285K. In the case of the $La^{3+}$-co-doped sample, three phases remain observable and the phase transitions are observed around 159K and 315 K. However, the comparison of the emission spectra of the $K_3Lu(PO_4)_2$:1%$Eu^{3+}$, 10%$La^{3+}$ observed below 150 K with the emission spectra of all phases of $K_3Lu(PO_4)_2$:1%$Eu^{3+}$ reveals several changes in the shape of the emission bands, which may suggest that for $K_3Lu(PO_4)_2$:1%$Eu^{3+}$, 10%$La^{3+}$ below 150 K co-existence of moth LT and MT phases are observed. These differences in the thermal stability of the individual structural phases can be primarily attributed to variations in the ionic radii of the introduced co-dopant ions (Figure 3e). Incorporation of these ions into the host lattice modifies the effective ionic-radius mismatch, expressed by the parameter Ω[17,28]:

$$\Omega = \frac{(1-x_i)R_0 + \mathrm{x}_i R_i}{R_0} \qquad (2)$$

where $R_0$ is the ionic radius of the host cation, $R_i$ is the ionic radius of the dopant ion, and $x_i$ denotes the molar fraction of the dopant. For a co-dopant concentration of 10%, Ω increases from 0.987 for $Sc^{3+}$, through 1.00553 for $Y^{3+}$, to 1.02086 for $La^{3+}$ (Figure 3f). In the case of $Sc^{3+}$, which possesses the smallest ionic radius among the investigated dopants, a significantly lower thermal energy is required to initiate the MT → HT phase transition. This observation suggests the presence of a less distorted polyhedral environment within the structure. Although this remains speculative, it is likely that the LT → MT transition in this composition occurs below the temperature range accessible in the present study. Alternatively, the absence of the LT phase may be related to the strong preference of $Sc^{3+}$ ions for six-fold coordination. Although the introduction of co-dopant ions significantly perturbs the thermodynamics of the phase transitions occurring between the two monoclinic phases, a remarkably systematic behavior is observed for the formation of the HT phase. The additional lattice strain introduced by the ionic-radius mismatch between the dopant and host ions leads to a monotonic variation of the HT phase-transition temperature ($T_{PT}$) (Figure 3g). Specifically, $T_{PT}$ increases from 224 K for 10%$Sc^{3+}$ to 324 K for 10%$La^{3+}$. This behavior demonstrates that co-doping provides an effective strategy for tailoring the phase-transition temperature and, consequently, offers a powerful tool for tuning the thermometric performance of luminescent thermometers based on $K_3Lu(PO_4)_2$:1%$Eu^{3+}$.

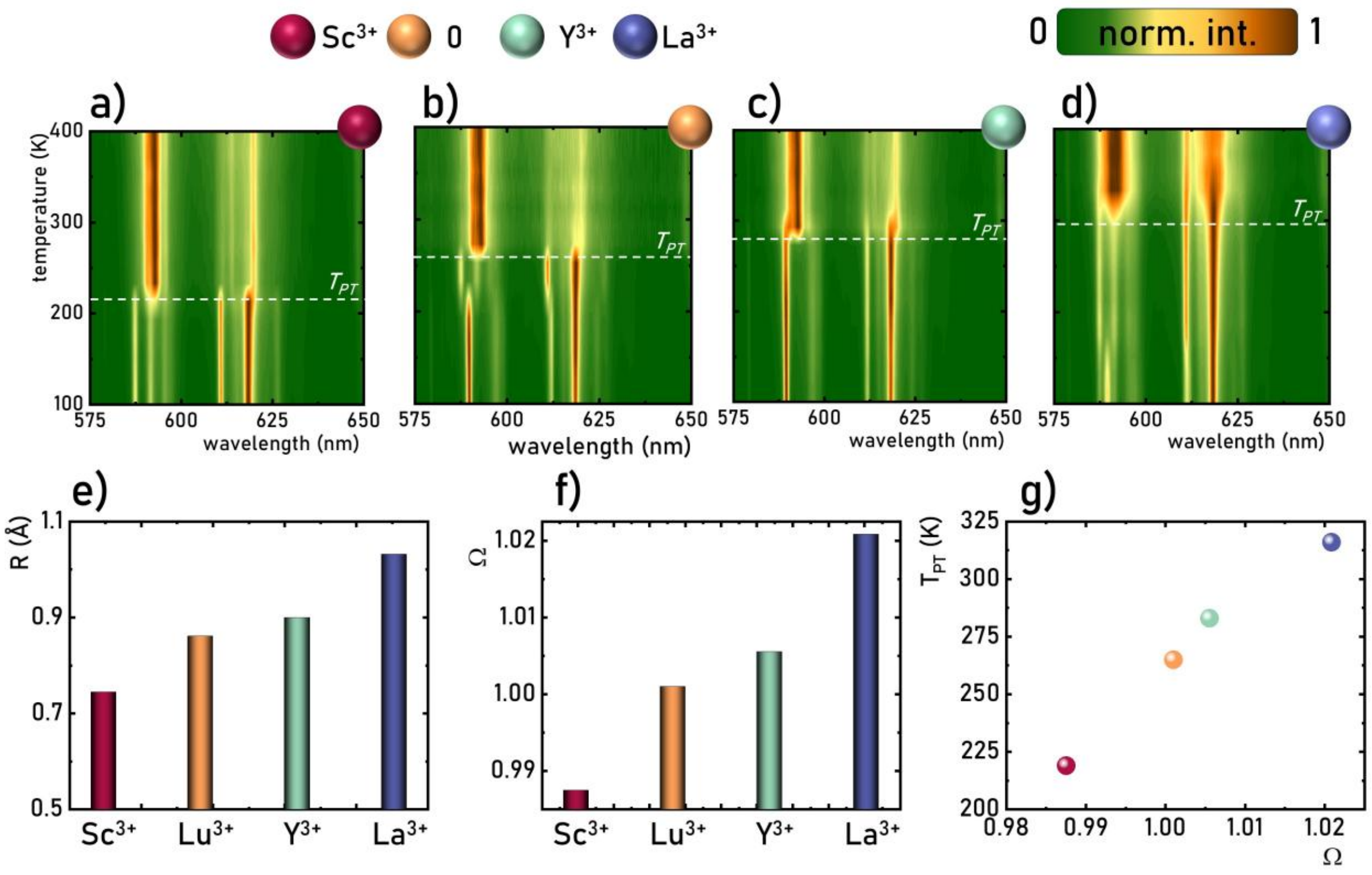

**Figure 3**. Thermal map of normalized emission spectra of $K_3Lu(PO_4)_2$:1%$Eu^{3+}$, 10% $Sc^{3+}$ - a); $K_3Lu(PO_4)_2$:1%$Eu^{3+}$ -b); $K_3Lu(PO_4)_2$:1%$Eu^{3+}$, 10% $Y^{3+}$ - c); $K_3Lu(PO_4)_2$:1%$Eu^{3+}$, 10% $La^{3+}$ - d); the comparison of ionic radii of co-dopants with $Lu^{3+}$ ions – e); the $\Omega$ for different ions – f), the $T_{PT}$ as a function of $\Omega$ – g).

To independently verify the effect of co-doping on the thermodynamics of the structural phase transitions, differential scanning calorimetry (DSC) measurements were performed for the investigated compositions upon both heating and cooling (Figure 4a). In all samples, the DSC traces reveal reversible thermal anomalies, confirming the occurrence of well-defined phase transformations and allowing their transition temperatures, enthalpy changes, and entropy changes to be determined. Although a small thermal hysteresis between heating and cooling is observed, the overall trends are highly consistent for both thermal cycles. A pronounced systematic dependence of the phase-transition temperature $T_{PT}$ on the ionic-radius mismatch parameter Ω is observed (Figure 4b). Increasing Ω results in a progressive increase in $T_{PT}$, from approximately 212 K for the $Sc^{3+}$-co-doped sample to 290 K for the $La^{3+}$-co-doped material, with the undoped reference and $Y^{3+}$-co-doped samples occupying intermediate positions. Thus, a relatively small variation in Ω produces a substantial shift of the transition temperature over a temperature interval exceeding 70 K. The same trend is reproduced upon heating and cooling, indicating that the correlation is not associated with a particular thermal history.

Importantly, the effect of Ω is not limited to the transition temperature but extends to the thermodynamic parameters of the transformation. The transition enthalpy ΔH increases systematically with increasing Ω, from approximately 1.4 kJ $mol^{-1}$ for the $Sc^{3+}$-co-doped sample to ca. 4.6 kJ $mol^{-1}$ for the $La^{3+}$-co-doped composition (Figure 4c). A similar monotonic evolution is observed for the transition entropy, which increases from approximately 7 to 16 J $mol^{-1}$ $K^{-1}$ over the investigated Ω range (Figure 4d). The close agreement between the values obtained upon heating and cooling further confirms the reversible character of the transformations. The simultaneous increase of $T_{PT}$, $\Delta H$, and $\Delta S$ with $\Omega$ provides an important thermodynamic perspective on the role of co-doping. Rather than simply shifting the temperature at which the structural transformation occurs, the ionic-radius mismatch systematically modifies both the energetic and configurational contributions associated with the transition. This behavior indicates that $\Omega$ can be considered a useful structural descriptor linking the magnitude of lattice perturbation introduced by co-doping with the thermodynamics of the phase transformation.

Taken together with the temperature-dependent luminescence results, the DSC measurements demonstrate that compositional modification provides a systematic means of controlling not only the thermal stability of the individual structural phases but also the thermodynamic driving forces associated with their transformation. This relationship is particularly relevant for luminescence thermometry, where the position and width of the phase-transition region determine the temperature range and sensitivity of the optical response.

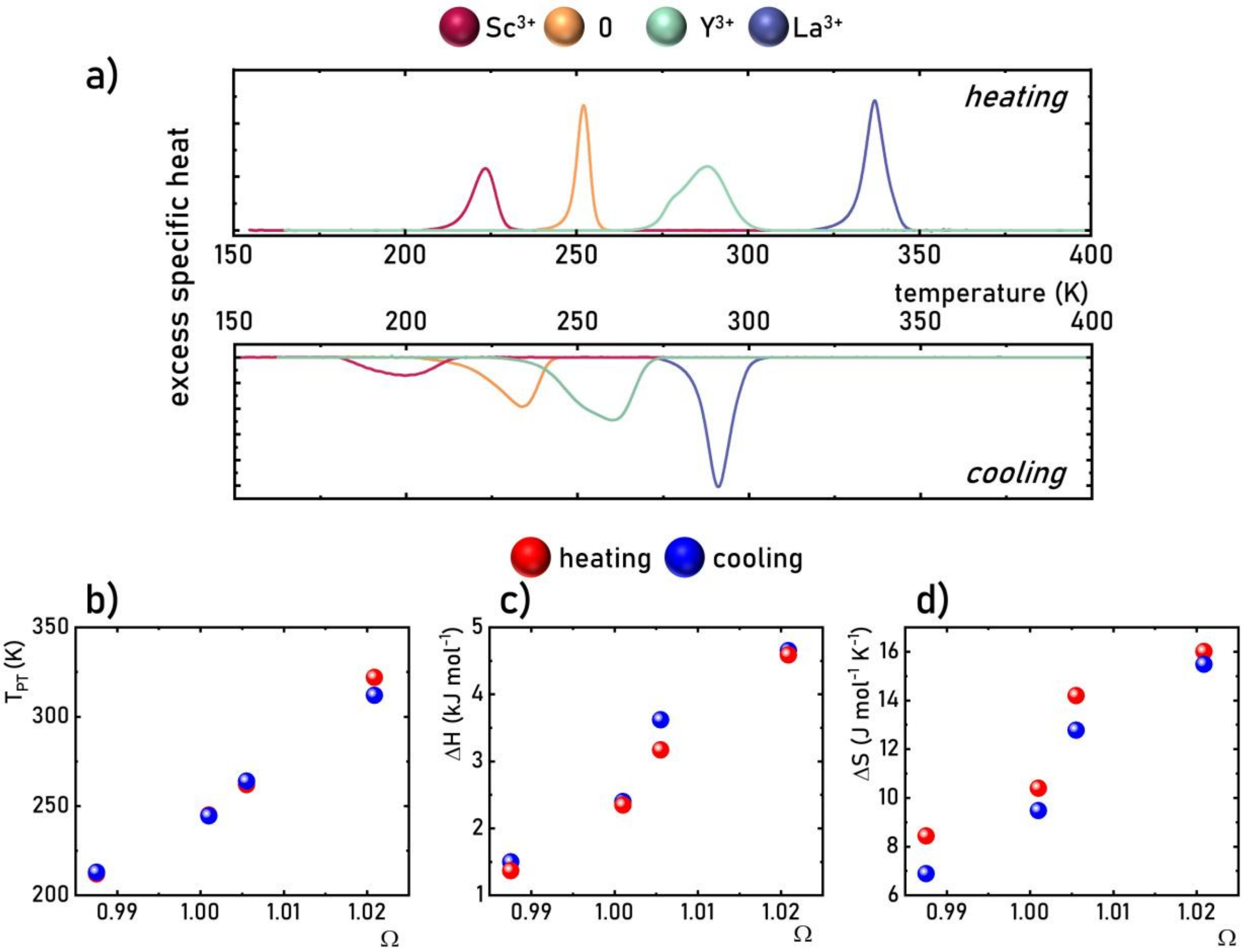

**Figure 4**. Correlation between structural distortion and the thermodynamics of the phase transition in $K_3Lu(PO_4)_2:Eu^{3+},RE^{3+}$- a); Excess heat-capacity profiles measured upon heating and cooling, respectively – b) Phase-transition temperature determined from the onset point ($T_{PT}$), -c) transition enthalpy ($\Delta H$), and - d) transition entropy ($\Delta S$) plotted as a function of the distortion parameter Ω.

As demonstrated above, the introduction of co-dopant ions such as $Sc^{3+}$, $La^{3+}$, or $Y^{3+}$ results in the modification of the thermal range in which MT phase is observed. This effect is associated with the difference in ionic radii between the host lattice cation and the introduced dopant ions. To verify this hypothesis, the influence of $Eu^{3+}$ concentration on the phase-transition temperatures $T_{PT1}$ and $T_{PT2}$ was investigated. Owing to the significantly smaller ionic-radius mismatch between $Eu^{3+}$ and $Lu^{3+}$, $Eu^{3+}$ doping provides a more suitable system for a systematic and precise analysis of the relationship between lattice strain and phase-transition behavior. For this purpose, temperature-dependent emission spectra of $K_3Lu(PO_4)_2$ doped with 0.1% $Eu^{3+}$, 1% $Eu^{3+}$, and 5% $Eu^{3+}$ were measured (Figure 5a-c, Figure S5, S6). Analysis of the thermal maps of the normalized emission spectra revealed that both phase transitions, namely LT → MT at $T_{PT1}$ and MT → HT at $T_{PT2}$, are clearly observed for all investigated compositions. However, a pronounced increase in both $T_{PT1}$ and $T_{PT2}$ is observed with increasing $Eu^{3+}$ concentration. These spectroscopic observations are in excellent agreement with the results obtained from differential scanning calorimetry (Figure 5d). A more detailed analysis shows that increasing the $Eu^{3+}$ concentration from 0.1% to 5% leads to an increase of $T_{PT1}$ from 150 K to 230 K, while $T_{PT2}$ shifts from 249 K to 290 K (Figure 5e). This behavior can be attributed to the increasing lattice strain induced by the incorporation of ions with ionic radii different from those of the host cations, an effect that has been widely reported in the literature. Importantly, increasing the $Eu^{3+}$ concentration not only shifts the phase-transition temperatures but also significantly narrows the temperature range over which the MT phase remains stable. To quantify this effect, the parameter Δ was defined as follows:

$$\Delta = T_{PT2} - T_{PT1} \qquad (3)$$

For $K_3Lu(PO_4)_2$:1%$Eu^{3+}$, increasing the $Eu^{3+}$ concentration from 0.1% to 5% reduces Δ from 95 K to 54 K (Figure 5f). At the same time, this increase in dopant concentration results in only a moderate increase of Ω. Considering the substantially larger Ω values obtained upon the introduction of 10% $Sc^{3+}$, $Y^{3+}$, or $La^{3+}$, a strong narrowing of the stability range of the MT phase, ultimately leading to its complete disappearance, is fully consistent with the proposed mechanism.

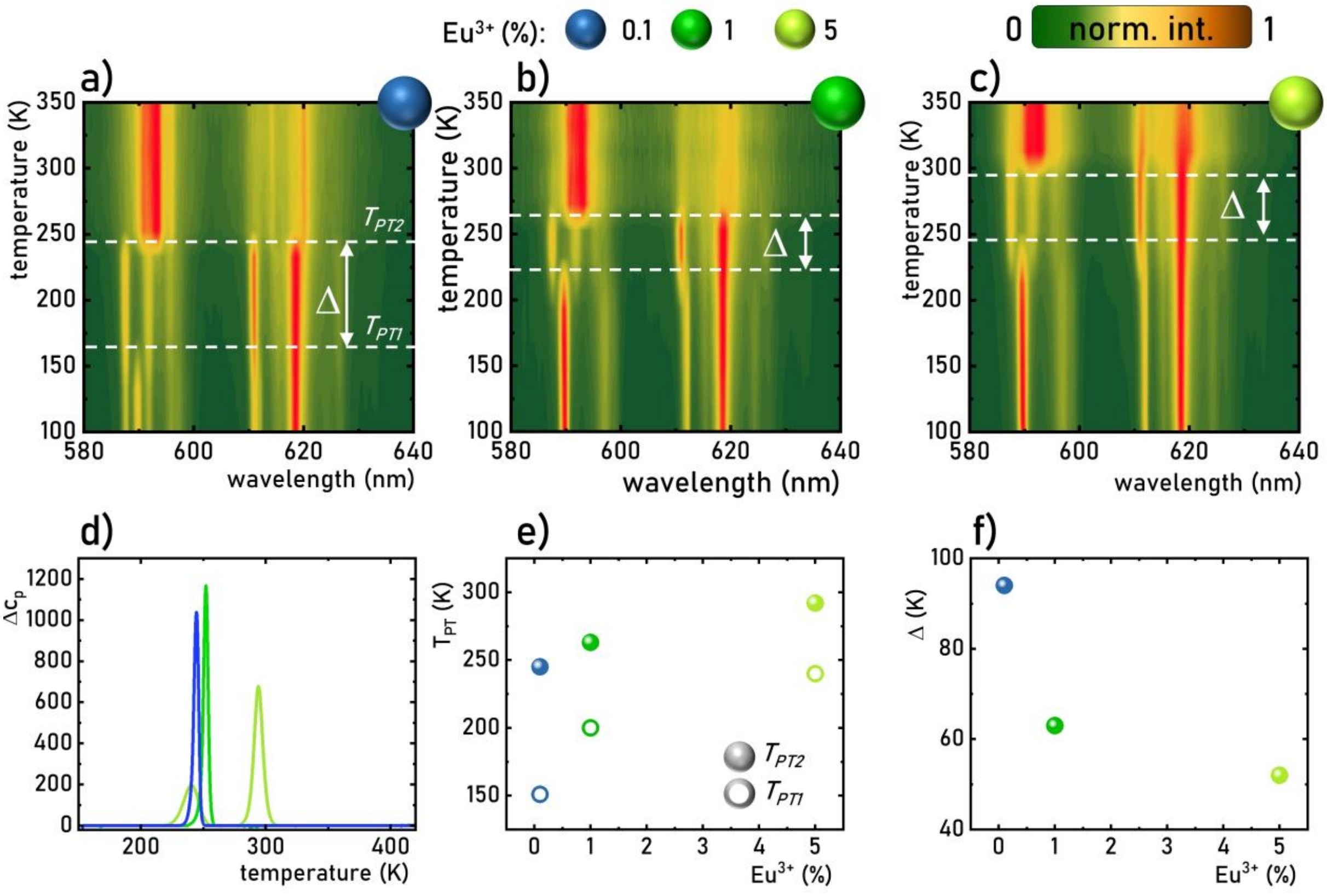

**Figure 5**. Thermal map of normalized emission spectra of $K_3Lu(PO_4)_2$:0.1%$Eu^{3+}$ -a); $K_3Lu(PO_4)_2$:1%$Eu^{3+}$ - b); $K_3Lu(PO_4)_2$:5%$Eu^{3+}$ - c); DSC curves for these compounds – d); the influence of $Eu^{3+}$ concentration on the $T_{PT1}$ and $T_{PT2}$ – e); the influence of $Eu^{3+}$ on the $\Delta$ – f).

The pronounced differences between the emission spectra observed for the MT and HT phases of $K_3Lu(PO_4)_2$:0.1%$Eu^{3+}$ provide an excellent opportunity for the development of a ratiometric luminescence thermometer. As demonstrated above, the structural phase transition in $K_3Lu(PO_4)_2$:0.1%$Eu^{3+}$ affects the intensity ratio between the $^5D_0 \rightarrow {}^7F_2$ and $^5D_0 \rightarrow {}^7F_1$ transitions of $Eu^{3+}$, defined as $LIR_1$. However, this parameter does not fully exploit the thermometric advantages associated with the phase transition. This limitation arises because both emission bands are present in the spectra of both structural phases, while only their relative intensities change with temperature. A significantly greater signal contrast is expected when spectral regions associated with Stark components characteristic of individual phases are used. Although changes in both the number and spectral positions of Stark lines are expected for all $Eu^{3+}$ emission transitions, the probability of spectral overlap between the MT- and HT-related signals increases with increasing $J$ value, since the number of Stark components arising from the splitting of the $^7F_J$ multiplets also increases. Consequently, the present analysis focuses on the $^5D_0 \rightarrow {}^7F_1$ transition, and the following thermometric parameter was proposed:

$$LIR_2 = \frac{\int_{591nm}^{593nm} {}^5D_0 \rightarrow {}^7F_1 \, d\lambda}{\int_{617nm}^{619nm} {}^5D_0 \rightarrow {}^7F_2 \, d\lambda} \quad (4)$$

Regardless of the co-dopant type, all investigated $K_3Lu(PO_4)_2$:0.1%$Eu^{3+}$-based materials exhibit a similar temperature dependence of $LIR_2$ (Figure 6a). At low temperatures, $LIR_2$ remains nearly constant and shows negligible temperature sensitivity until a threshold temperature corresponding to the structural phase transition is reached. At this point, $LIR_2$ increases abruptly within a narrow temperature interval. Further temperature increase leads to saturation of the parameter. The temperature at which this transition occurs, and consequently the thermal operating range of the thermometer, strongly depends on the nature of the co-dopant ion. To quantitatively describe the thermometric performance, the relative sensitivity ($S_R$) was calculated according to:

$$S_R = \frac{1}{LIR} \frac{\Delta LIR}{\Delta T} \cdot 100\% \quad (5)$$

The rapid variation of $LIR_2$ within a narrow temperature range results in high $S_R$ values only in the immediate vicinity of the phase-transition temperature (Figure 6b). The maximum relative sensitivities obtained for the investigated phosphors are $S_{Rmax}$ = 18.8% $K^{-1}$ at 253 K for $K_3Lu(PO_4)_2$:1%$Eu^{3+}$, 14.6% $K^{-1}$ at 204 K for $K_3Lu(PO_4)_2$:1%$Eu^{3+}$,10%$Sc^{3+}$, 6.7% $K^{-1}$ at 273 K for $K_3Lu(PO_4)_2$:1%$Eu^{3+}$, 10%$Y^{3+}$, and 3.1% $K^{-1}$ at 299 K for $K_3Lu(PO_4)_2$:1%$Eu^{3+}$, 10% $La^{3+}$. These results clearly demonstrate that co-doping not only modifies the temperature at which the maximum sensitivity occurs but also significantly affects the magnitude of $S_{Rmax}$. The monotonic variation of the temperature corresponding to $S_{Rmax}$ ($T@S_{Rmax}$) with the Ω parameter indicates that the thermometric properties of the material can be systematically tailored through compositional engineering (Figure 6c). It is also noteworthy that, although co-doping reduces $S_{Rmax}$, it simultaneously broadens the full width at half maximum of the $S_R$ thermal dependence. This effect has important implications for the thermal operating range of the luminescence thermometer. According to the commonly accepted criterion in the literature, the thermal operating range is defined as the temperature interval over which the thermometric parameter exhibits a monotonic temperature dependence and $S_R$ exceeds 1% $K^{-1}$ [41]. Based on this definition, the thermal operating range shifts toward higher temperatures and becomes progressively broader with increasing Ω, namely: 191-221 K for $K_3Lu(PO_4)_2$:1%$Eu^{3+}$, 10% $Sc^{3+}$, 229-265 K for $K_3Lu(PO_4)_2$:1%$Eu^{3+}$, 238-306 K for $K_3Lu(PO_4)_2$:1%$Eu^{3+}$, 10% $Y^{3+}$, and 268-328 K for $K_3Lu(PO_4)_2$:1%$Eu^{3+}$, 10% $La^{3+}$ (Figure 6d). From an application perspective, this observation is particularly important. While maximizing $S_R$ is often desirable, many practical applications require a broader operating temperature range, and a relative sensitivity of approximately 1% $K^{-1}$ is already sufficient for reliable temperature determination. Under such circumstances, co-dopants that induce larger changes in Ω may be preferred, as they provide a wider and more versatile thermal operating window while maintaining satisfactory thermometric performance.

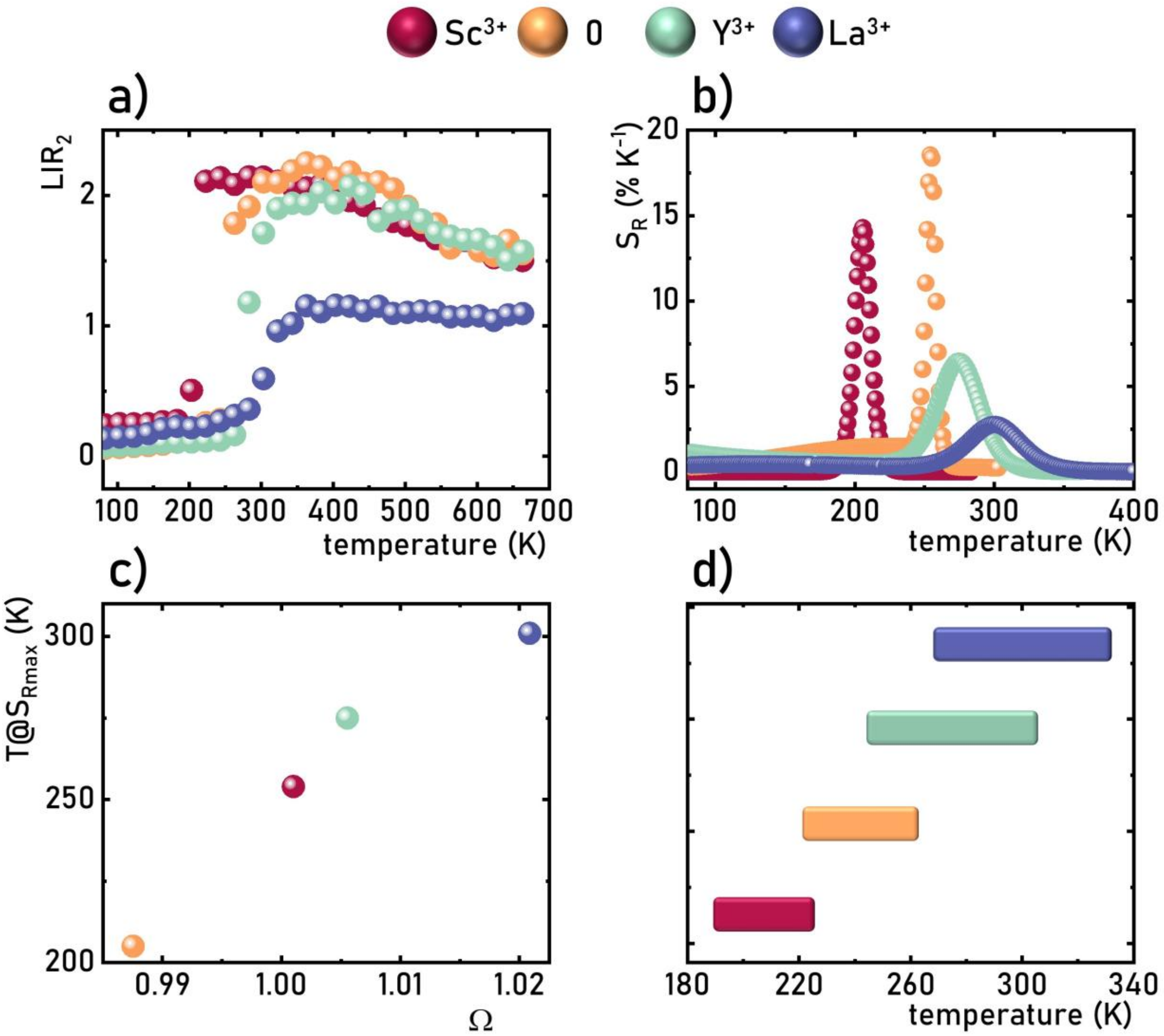

**Figure 6**. Thermal dependence of $LIR_1$ for different phosphors – a) and corresponding $S_R$ – b); $T@S_{Rmax}$ as a function of $\Omega$ – c); thermal operating range of different luminescence thermometers– d).

The increase in $\Delta H$ and $\Delta S$ with increasing Ω can be attributed to progressively larger energetic and entropic changes associated with the phase transition, reflecting a greater difference in the structural and dynamic states of the two phases. Since the transition temperature is determined by the balance between these two contributions, according to $T_{PT}=\Delta H/\Delta S$, their different rates of increase also account for the systematic shift of the transition temperature with Ω. This effect is manifested not only in the broadening of the transition features observed in the DSC profiles, but also in the temperature dependence of the relative sensitivity, $S_R$, as demonstrated above. From a thermometric perspective, this behavior has important implications because it directly affects the usable thermal range of the luminescence thermometer. Analysis of the full width at half maximum (*FWHM*) of the $S_R(T)$ profiles reveals that the *FWHM* progressively increases with Ω for the $K_3Lu(PO_4)_2$:1%$Eu^{3+}$ series (Figure 7a). Importantly, an analogous analysis performed for previously reported phase-transition-based luminescence thermometers, including $LaGaO_3$:$Eu^{3+}$, $Al^{3+}$[28] and $LiYO_2$:$Eu^{3+}$, $Na^{+}$ [27], reveals similar linear trends (Figure 7b). Although the slopes of the *FWHM(Ω)* relationships differ among the investigated material systems, the same general tendency is consistently observed, indicating that an increasing ionic-radius mismatch is accompanied by a broadening of the temperature interval over which enhanced relative sensitivity is maintained. Nevertheless, the *FWHM* of the $S_R(T)$ profile does not provide a sufficiently representative measure of the practically usable temperature range of a luminescence thermometer. A more application-relevant criterion commonly adopted in luminescence thermometry defines the usable thermal range (*UTR*) as the temperature interval over which the thermometric parameter exhibits a monotonic temperature dependence while maintaining $S_R > 1\%\ K^{-1}$ (Figure 7a) [41]. According to this criterion, the *UTR* for the $K_3Lu(PO_4)_2$:1%$Eu^{3+}$ series increases substantially, from 32 K for $K_3Lu(PO_4)_2$:1%$Eu^{3+}$ to 60 K for $K_3Lu(PO_4)_2$:1%$Eu^{3+}$:10%$La^{3+}$. Moreover, the *UTR* exhibits a monotonic increase with increasing Ω (Figure 7c). Notably, analogous relationships are observed for the $LiYO_2$:$Eu^{3+}$ and $LaGaO_3$:$Eu^{3+}$ systems, further supporting the broader applicability of this trend. These results demonstrate that controlling Ω through the deliberate selection of co-dopant ions provides a means not only to tune the phase-transition temperature but also to broaden the usable thermal range of phase-transition-based luminescence thermometers. This finding is particularly important from an application perspective, as it establishes a straightforward compositional strategy for the intentional engineering of both the operating temperature and the useful temperature range of luminescence thermometers, enabling their thermometric performance to be tailored toward specific application requirements.

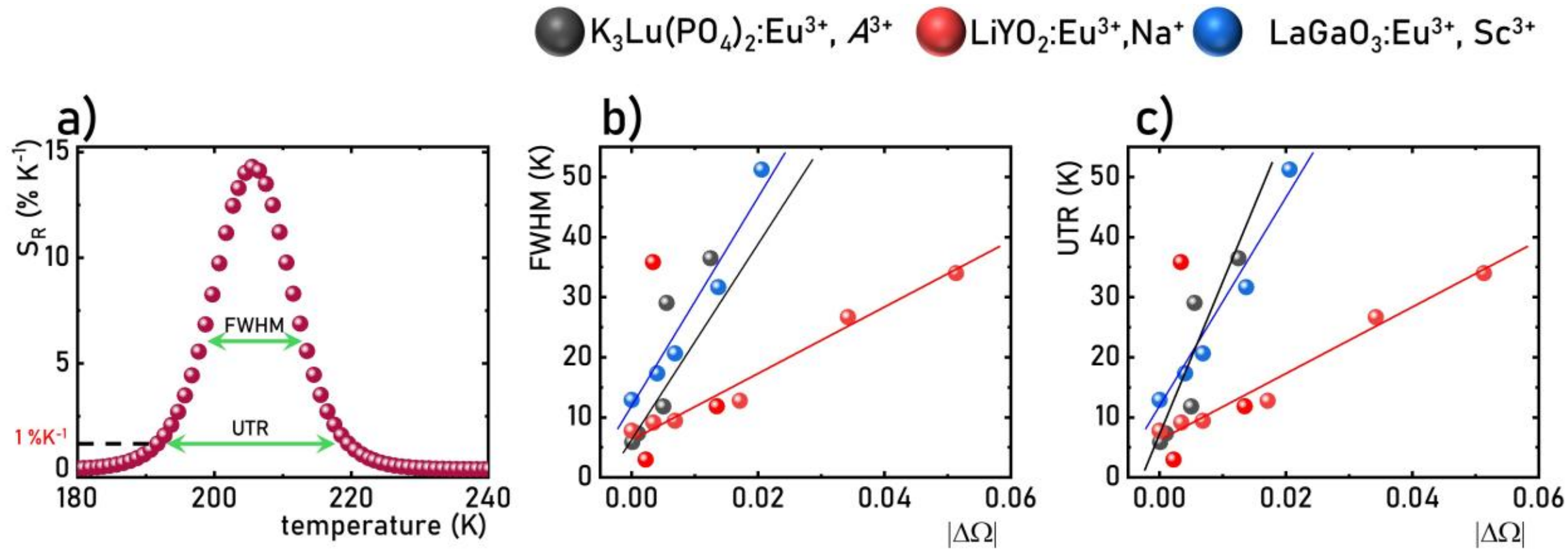

**Figure 7**. Schematic presentation of the *FWHM* and *UTR* on the thermal dependence of $S_R$ – a); the dependence of *FWHM* – b) and *UTR* – c) on |ΔΩ| for $K_3Lu(PO_4)_2:Eu^{3+}$, $A^{3+}$ (where *A*=La, Y, Sc), $LiYO_2:Eu^{3+},Na^+$ and $LaGaO_3:Eu^{3+},Sc^{3+}$.

## Conclusions

This work presents a systematic investigation of the temperature-dependent spectroscopic properties of $K_3Lu(PO_4)_2:Eu^{3+}$ aimed at elucidating the influence of co-dopant incorporation on the thermometric performance of a luminescence thermometer based on a first-order structural phase transition in this material. Two phase transitions, $\alpha \rightarrow \beta$ and $\beta \rightarrow \gamma$, were identified at approximately 230 and 270 K, respectively. These structural transformations modify the local point symmetry of the crystallographic sites occupied by $Eu^{3+}$ ions, thereby inducing pronounced changes in their spectroscopic properties. In particular, changes in the number of Stark components arising from the splitting of individual $Eu^{3+}$ energy levels across the phase transitions, and the resulting modification of the emission spectral profile, were exploited to develop a ratiometric luminescence thermometer. A maximum relative sensitivity of $S_{Rmax}$ = 18.8% K$^{-1}$ at 253 K was achieved for $K_3Lu(PO_4)_2$:1%$Eu^{3+}$. The incorporation of $Y^{3+}$, $La^{3+}$, and $Sc^{3+}$ co-dopants substantially modifies the thermometric properties of the system due to the differences between their ionic radii and that of the substituted $Lu^{3+}$ ions. Most notably, co-doping enables a pronounced modulation of the phase-transition temperature, ranging from 224 K for the sample containing 10%$Sc^{3+}$ to 324 K for that containing 10%$La^{3+}$. To quantitatively describe this effect, a relative ionic-radius mismatch parameter, Ω, was introduced, with values of 0.99 for $Sc^{3+}$, through 1.004 for $Y^{3+}$, to 1.018 for $La^{3+}$ for the respective compositions. Importantly, Ω provides a simple structural descriptor of the perturbation introduced by co-doping and serves as a practical tuning parameter linking the ionic-radius mismatch to the phase-transition temperature and, consequently, to the thermometric operating range. The phase-transition temperature, $T_{PT}$, determined from the temperature-dependent luminescence spectra was found to increase monotonically with increasing Ω across the investigated phosphor series. These results are consistent with those obtained from differential scanning calorimetry, which additionally revealed linear increases in both the transition enthalpy, $\Delta H$, and transition entropy, $\Delta S$, with increasing Ω. The controlled shift of the phase-transition temperature directly translates into a corresponding displacement of the thermal operating range of the luminescence thermometer. Moreover, the increase in both ΔH and ΔS induced by co-dopant incorporation reflects progressively larger energetic and entropic changes accompanying the structural transformation, as evidenced by the broadening of the corresponding DSC feature. Consequently, the temperature dependence of the relative sensitivity, $S_R$, also becomes broader. Although this effect leads to a reduction in the maximum achievable relative sensitivity, it simultaneously extends the temperature interval over which high thermometric performance can be maintained. Accordingly, both the full width at half maximum (*FWHM*) of the $S_R$ profile and the usable thermal operating range were found to increase sublinearly with increasing Ω. These results reveal an inherent trade-off between the maximum relative sensitivity and the width of the useful temperature range, which can be deliberately controlled through the ionic-radius mismatch introduced by co-doping. Importantly, comparative analysis indicates that this behavior is not specific to $K_3Lu(PO_4)_2:Eu^{3+}$. Similar relationships were also identified for $LaGaO_3:Eu^{3+}$ and $LiYO_2:Eu^{3+}$-based systems, suggesting that the proposed strategy for tailoring the performance of phase-transition-based luminescence thermometers, and, more generally, the characteristics of materials exhibiting structural phase transitions, may have broader applicability, at least within the investigated family of materials. Further studies involving a wider range of host lattices and structural phase-transition mechanisms are required to establish the generality of these relationships. Nevertheless, the results presented here provide a promising framework for the rational compositional engineering of phase-transition materials with deliberately tailored transition temperatures, thermodynamic parameters, and operating ranges, potentially enabling their optimization for a variety of applications.

**Acknowledgements**

This work was supported by the National Science Center (NCN) Poland under project no. DEC-UMO-2022/45/B/ST5/01629. M.Sz. gratefully acknowledges the support of the Foundation for Polish Science through the START program.